\documentclass[aps,prb,reprint,superscriptaddress]{revtex4-1}

\usepackage{graphicx}
\usepackage{siunitx}

\begin{document}


\title{Stacking domains in graphene on silicon carbide: a pathway for intercalation}

\author{T.A.\ de Jong}
\affiliation{Huygens-Kamerlingh Onnes Laboratorium, Leiden Institute of Physics, Leiden University, Niels Bohrweg 2, P.O. Box 9504, NL-2300 RA Leiden, The Netherlands}

\author{E.E.\ Krasovskii}
\affiliation{Departamento de F\'isica de Materiales, Universidad del Pais Vasco UPV/EHU, 20080 San Sebasti\'an/Donostia, Spain}
\affiliation{IKERBASQUE, Basque Foundation for Science, E-48013 Bilbao, Spain}
\affiliation{Donostia International Physics Center (DIPC), E-20018 San Sebasti\'an, Spain}

\author{C.\ Ott}
\affiliation{Lehrstuhl f\"ur Angewandte Physik, Universit\"at Erlangen-N\"urnberg, 91058 Erlangen, Germany}

\author{R.M.\ Tromp}
\affiliation{IBM T.J.Watson Research Center, 1101 Kitchawan Road, P.O.\ Box 218, Yorktown Heights, New York, New York 10598, USA}
\affiliation{Huygens-Kamerlingh Onnes Laboratorium, Leiden Institute of Physics, Leiden University, Niels Bohrweg 2, P.O. Box 9504, NL-2300 RA Leiden, The Netherlands}

\author{S.J.\ van der Molen}
\affiliation{Huygens-Kamerlingh Onnes Laboratorium, Leiden Institute of Physics, Leiden University, Niels Bohrweg 2, P.O. Box 9504, NL-2300 RA Leiden, The Netherlands}

\author{J.\ Jobst}
\email{jobst@physics.leidenuniv.nl}
\affiliation{Huygens-Kamerlingh Onnes Laboratorium, Leiden Institute of Physics, Leiden University, Niels Bohrweg 2, P.O. Box 9504, NL-2300 RA Leiden, The Netherlands}

\date{\today}

\begin{abstract}
Graphene on silicon carbide (SiC) bears great potential for future graphene electronic applications \cite{gaskill2009-power-electronics, avouris-GHz, kubatkin-QHE, hertel-monolithic, bianco2015-THz} because it is available on the wafer-scale \cite{emtsev2009-towards, deHeer2011-confined-growth, lin-waferscale} and its properties can be custom-tailored by inserting various atoms into the graphene/SiC interface \cite{riedl2009quasi, emtsev-Ge-intercalation, nandkishore2012-SC-doping, li2013-TI-intercalation, baringhaus2015-ballistic-Ge, anderson2017-Eu-intercalation, speck2017growth}. It remains unclear, however, how atoms can cross the impermeable graphene layer during this widely used intercalation process \cite{riedl2009quasi, berry2013-impermeability, Hu2014-proton-transport}. Here we demonstrate that, in contrast to the current consensus, graphene layers on SiC are not homogeneous, but instead composed of domains of different crystallographic stacking  \cite{hibino2009stacking, alden2013strain, butz2014dislocations}. We show that these domains are intrinsically formed during growth and that dislocations between domains dominate the (de)intercalation dynamics. Tailoring these dislocation networks, e.g. through substrate engineering, will increase the control over the intercalation process and could open a playground for topological and correlated electron phenomena in two-dimensional superstructures \cite{ju2015-TI-transport,hunt2013-butterfly,herrero2018-magic-Mott,herrero2018-superconductivity}.
\end{abstract}


\maketitle

Graphene can routinely be produced on the wafer scale by thermal decomposition of silicon carbide (SiC) \cite{emtsev2009-towards, deHeer2011-confined-growth, lin-waferscale}. Due to the direct growth on SiC(0001) wafers, epitaxial graphene (EG) naturally forms on a wide band gap semiconductor, providing a doped or insulating substrate compatible with standard CMOS fabrication methods. Hence, EG is a contender for future graphene electronic applications such as power electronics \cite{gaskill2009-power-electronics, hertel-monolithic}, high-speed transistors \cite{avouris-GHz}, quantum resistance standards \cite{kubatkin-QHE} and terahertz detection \cite{bianco2015-THz}.
In EG, the first hexagonal graphene layer resides on an electrically insulating monolayer of carbon atoms that are sp$^3$ bonded to silicon atoms of the SiC(0001) surface \cite{emtsev2009-towards, deHeer2011-confined-growth, lin-waferscale, tanaka2010anisotropic}.
The presence of this so-called buffer layer strongly affects the graphene on top, e.g. by pinning the Fermi level. 
Consequently, the graphene properties can be tuned via intercalation of atoms into the buffer layer/SiC interface. 
The intercalation of hydrogen is most widely used and results in the conversion of the buffer layer to a quasi-freestanding graphene (QFG) layer by cutting the silicon-carbon bonds and saturating silicon dangling bonds with hydrogen. 
This treatment reverses the graphene doping from n-type to p-type and improves the mobility \cite{riedl2009quasi, speck-QFMLG}. 
Intercalation of heavier atoms is used to further tailor the graphene properties, e.g. to form pn-junctions \cite{emtsev-Ge-intercalation, baringhaus2015-ballistic-Ge}, magnetic moments \cite{anderson2017-Eu-intercalation} or potentially superconducting \cite{nandkishore2012-SC-doping} and topologically non-trivial states \cite{li2013-TI-intercalation}.

\begin{figure*}[t]
	\centering  
	\includegraphics[width=0.9\textwidth]{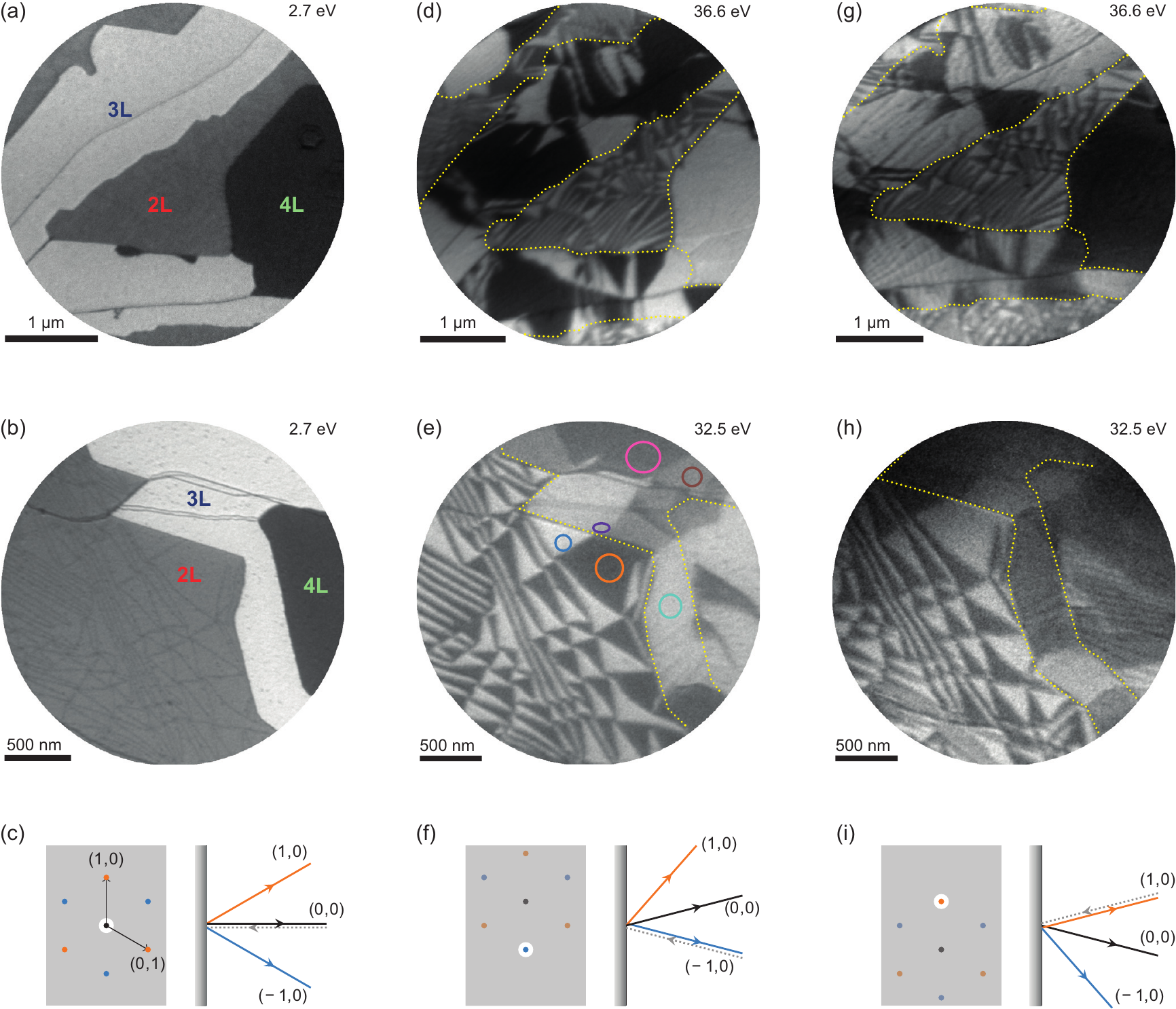}
	\caption{Graphene on SiC is composed of domains of different stacking order.
		(a, b) Bright-field LEEM micrographs of two samples of bilayer, trilayer and four-layer QFG. 
		(c) In bright-field geometry, images are recorded from specularly reflected electrons (black) by selecting the $(0,\!0)$ diffraction spot using an aperture (gray rectangle) that blocks all diffracted beams (orange and blue). 
		(d, e) Dark-field images of the same area as in a,b. Domains of alternating contrast are clearly visible, indicating areas of different stacking order.
		(f) Sketch of the tilted dark-field geometry selecting the $(-1,\!0)$ spot as used for d,e. 
		(g, h) Dark-field images using the inequivalent $(1,\!0)$ diffraction spot show inverted contrast compared to d,e.
		(i) Measurement geometry used for g,h. See Methods for details on LEEM imaging modes.
		Yellow lines in d,e,g,h are guides to the eye indicating areas of constant layer number. Circles in e indicate areas from which the spectroscopy data in Fig.\ \ref{fig:spectra}c,d is obtained.}\label{fig:domains}
\end{figure*}

Graphene on SiC (EG and QFG) appears homogeneous with low defect concentration in most techniques \cite{emtsev2009-towards, deHeer2011-confined-growth, lin-waferscale, riedl2009quasi}. Together with the fact that layers span virtually unperturbed over SiC substrate steps \cite{lauffer-STM, ross-steps, kautz2015-LEEP}, this has led to the consensus of perfectly crystalline graphene.
On the other hand, two observations point to a less perfect sheet. 
First, the charge carrier mobility is generally low, even at cryogenic temperatures  \cite{speck-QFMLG, emtsev2009-towards}. 
Second, an ideal graphene sheet is impermeable even to hydrogen \cite{berry2013-impermeability, Hu2014-proton-transport}, whereas a wide variety of atomic and molecular species has been intercalated into EG \cite{riedl2009quasi, nandkishore2012-SC-doping, li2013-TI-intercalation, baringhaus2015-ballistic-Ge, anderson2017-Eu-intercalation, speck2017growth}.
In this Report, we demonstrate that graphene on SiC is less homogeneous than widely believed and is, in fact, fractured into domains of different crystallographic stacking order. 
We use advanced low-energy electron microscopy (LEEM) methods and \emph{ab initio} calculations to show that those domains are naturally formed during growth due to nucleation dynamics and built-in strain. They are thus present in all graphene-on-SiC materials.

Figure \ref{fig:domains}(a) and (b) show bright-field LEEM images of two QFG samples (see Methods section for details on sample growth and hydrogen intercalation) with areas of different graphene thickness. 
Bright-field images are recorded using specularly reflected electrons that leave the sample perpendicular to the surface (see Fig.\ \ref{fig:domains}(c)). 
The main contrast mechanism in this mode is the interaction of the imaging electrons with the thickness-dependent, unoccupied band structure of the material, which is used to unambiguously determine the number of graphene layers \cite{hibino2008-thickness, feenstra2013-QFG-IV, jobst2015-ARRES}.
Large, homogeneous areas of bilayer, trilayer and four-layer graphene can thus be distinguished in Fig.\ \ref{fig:domains}(a,b), supporting the notion of perfect crystallinity.  

In stark opposition to this generally accepted view, the dark-field images in Fig.\ \ref{fig:domains}(d,e) clearly reveal that all areas are actually fractured into domains of alternating contrast.
The symmetry breaking introduced in dark-field imaging, where the image is formed from one diffracted beam only (cf.\ Fig.\ \ref{fig:domains}(f) and Methods), leads to strong contrast between different stacking types of the graphene layers \cite{hibino2009stacking, speck2017growth}. 
In fact, the contrast between different domains inverts (Fig.\ \ref{fig:domains}(d,e) versus (g,h)) when dark-field images are recorded from non-equivalent diffracted beams (cf.\ Fig.\ \ref{fig:domains}(f) and (i)).

\begin{figure}[!ht]
	\centering
	\includegraphics[width=\columnwidth]{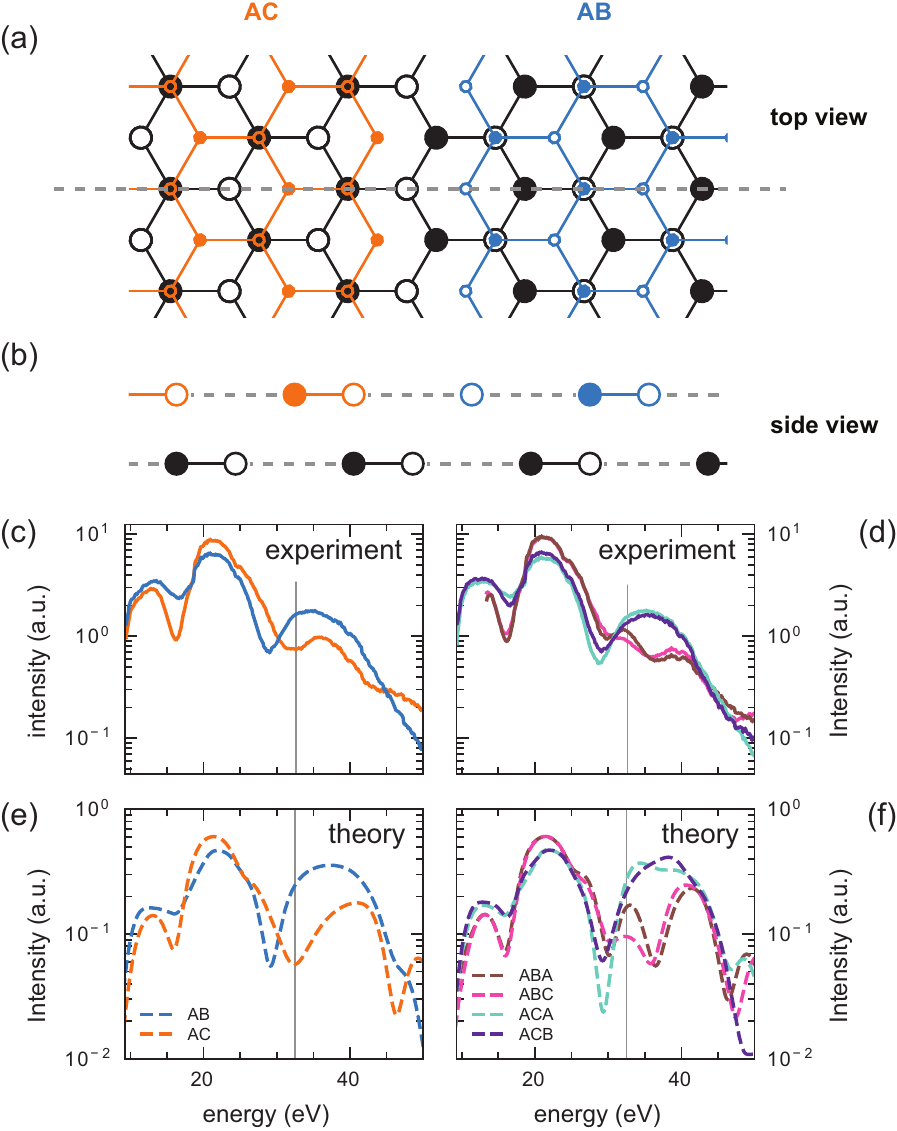}
	\caption{Low-energy electron reflectivity spectra reveal precise stacking order.
		(a) Sketched top view of AC (orange) and AB (blue) stacking orders. Inequivalent atoms of the unit cell of the top layer (orange or blue) sit in the center of the hexagon of the bottom layer (black). 
		(b) Side view of the stacking along the dashed line in A. Open and closed circles denote the inequivalent atoms of the graphene unit cell.
		(c, d) Experimental dark-field reflectivity spectra recorded on different stacking domains on bilayer and trilayer graphene, respectively. The areas from which the spectra are recorded are indicated by circles in Fig.\ \ref{fig:domains}(e).
		(e, f) Theoretical dark-field spectra for AB and AC as well as ABA, ABC, ACA and ACB stacking orders obtained by \emph{ab initio} calculations. A Gaussian broadening of \SI{1}{\electronvolt} is applied to account for experimental losses. 
		The vertical lines in (c) to (f) indicate the landing energy at which Fig.\ \ref{fig:domains}(e,h) are recorded. }
	\label{fig:spectra}
\end{figure}

At first glance, the observation of different stacking orders is surprising, as it is known that graphene layers grown on SiC(0001) are arranged in Bernal stacking \cite{deHeer2011-confined-growth,hibino2009stacking}. 
However, two energetically equivalent versions of Bernal stacking exist, AB and AC.   
The AC stacking order can be thought of either as AB bilayer where the top layer is translated by one bond length, or alternatively, as a full AB bilayer rotated by 60 degrees (Fig.\ \ref{fig:spectra}(a,b)). Consequently, AB and AC stacking are indistinguishable in bright-field imaging. 
Subsequent layers can be added in either orientation, generating more complicated stacking orders for trilayer and beyond. 

In order to identify the exact stacking in each area, we simulate bilayer and trilayer graphene slabs in different stacking orders and compare their reflectivity with measured low-energy electron reflectivity spectra. 
The latter are extracted from the intensity of an area in a series of spectroscopic LEEM images recorded at different electron landing energy (see Supplementary Movie 1 and 2 for such measurements of the area in Fig.\ \ref{fig:domains}(b) in bright-field and dark-field geometry, respectively). 
While different domains show identical bright-field reflectivity (cf.\ Supplementary Figure 1), 
dark-field spectra extracted from different bilayer domains (marked blue and orange in Fig.\ \ref{fig:spectra}(c) and \ref{fig:domains}(e) are clearly distinguishable. Moreover, four distinct reflectivity curves are observed for trilayer graphene (Fig.\ \ref{fig:spectra}(d)).
Figure \ref{fig:spectra}(e,f) shows theoretical dark-field spectra, obtained by \emph{ab initio} calculations (see Methods section for computational details), of different bilayer and trilayer stacking orders, respectively.
The excellent agreement of theoretical and experimental data in Fig.\ \ref{fig:spectra}(c,e) is clear evidence that the assignment of Bernal AB and AC stacking orders for different bilayer domains is correct. 
Moreover, the comparison of Fig.\ \ref{fig:spectra}(d) and (f) shows that using these dark-field LEEM methods, we can distinguish the more complicated trilayer stacking orders: Bernal, ABA (cyan) and ACA (pink), versus rhombohedral ABC (purple) and ACB (brown).
Due to the small electron penetration depth in LEEM, however, the spectra fall into two families (ABA and ABC vs.\ ACA and ACB) dominated by the stacking order of the top two layers.

\begin{figure}[t]
	\centering
	\includegraphics[width=\columnwidth]{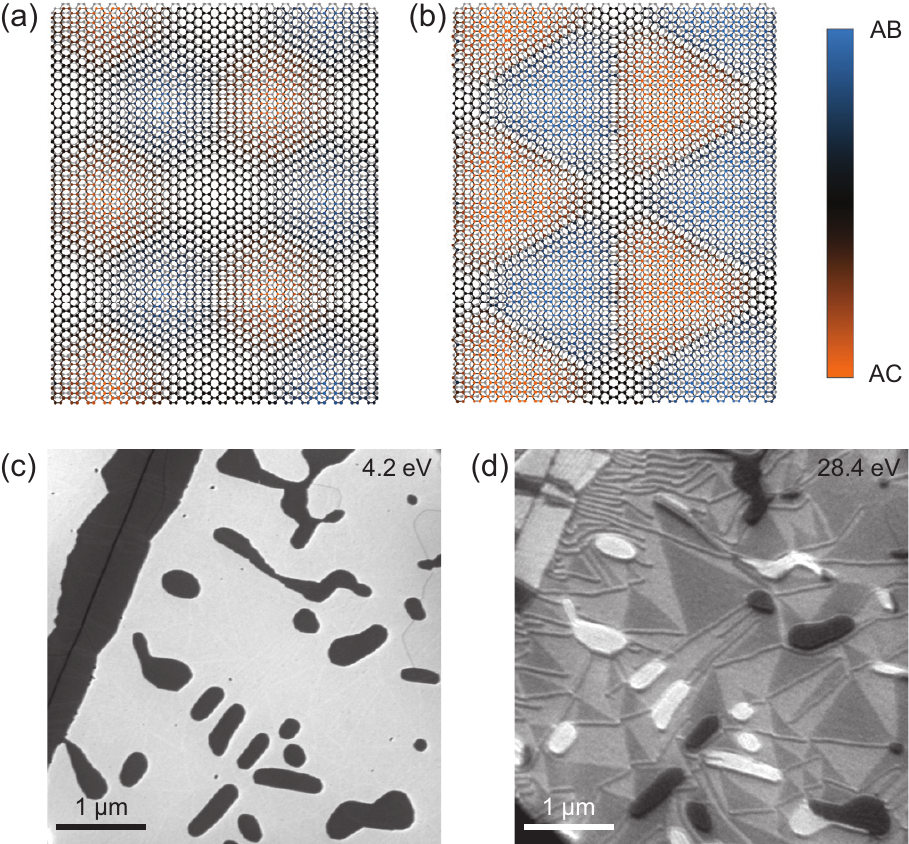}
	\caption{Stacking domains are caused by growth-induced strain and graphene nucleation dynamics.
		(a) Sketch of bilayer graphene where the top layer is uniformly strained causing a Moir\'e pattern. 
		(b) Sketch of the energetically favored arrangement of AB and AC stacked domains with all strain concentrated into dislocation lines. The trigonal shape of the domains is clearly visible. The color denotes how close a local stacking order is to AB (orange) or AC (blue) stacking.
		(c) A bright-field LEEM image of EG where growth was stopped shortly after bilayer starts to form. 
		(d) Dark-field LEEM of the same area reveals that the resulting islands, which emerged from individual nucleation sites, exhibit constant stacking order, i.e. they are either AB (bright) or AC (dark) stacked. }
	\label{fig:nucleation}
\end{figure}

In addition to their stacking orders, bilayer graphene and thicker areas differ in the morphology of the stacking domains (cf.\ Fig.\ \ref{fig:domains}(d,e)), which indicates two distinct formation mechanisms.
Most notably, bilayer domains are smaller, triangular and relatively regular. 
Similar morphologies, observed in free-standing bilayer graphene~\cite{butz2014dislocations} and graphene grown on copper~\cite{Brown2012,alden2013strain}, were linked to strain between the layers. 
While uniform strain causes a Moir\'e reconstruction (Fig.\ \ref{fig:nucleation}(a)), it is often energetically favorable to form domains of commensurate, optimal Bernal stacking. 
In this case, all strain is concentrated into the domain walls, thus forming dislocation lines~\cite{butz2014dislocations,alden2013strain}, as sketched in Fig.\ \ref{fig:nucleation}(b). 
Upon close examination of Fig.\ \ref{fig:domains}(b), the network of these dislocations is visible as dark lines in our bright-field measurements.
The size of the triangular domains shrinks for increasing uniform strain, while anisotropic strain causes domains elongated perpendicular to the strain axis. 
The observed average domain diameter of $\sim$100--200\,nm coincides well with relaxation of the 0.2\% lattice mismatch between buffer layer and first graphene layer~\cite{schumann2014effect} (see calculation in the Supplementary Information).
We thus conclude that the triangular domains in bilayer graphene result from strain thermally induced during growth and from the lattice mismatch with the SiC substrate. 
The presence of elongated triangular domains indicates non-uniform strain due to pinning to defects and substrate steps.

The larger, irregularly shaped domains that dominate trilayer and four-layer areas (Fig.\ \ref{fig:domains}(d,g)) can be explained by nucleation kinetics. 
To test this hypothesis, we study EG samples where the growth was stopped shortly after the nucleation of bilayer areas to prevent their coalescence (see Methods). The resulting small bilayer islands on monolayer terraces are shown in bright-field and dark-field conditions in Fig.\ \ref{fig:nucleation}(c) and (d), respectively. 
We observe that bilayer areas with a diameter below $\sim$300\,nm form single domains of constant stacking order (either bright or dark in Fig.\ \ref{fig:nucleation}(d)) and that AB and AC stacked bilayer islands occur in roughly equal number.
This indicates that new layers nucleate below existing ones in one of the two Bernal stacking orders randomly \cite{emtsev2009-towards, deHeer2011-confined-growth, lin-waferscale, tanaka2010anisotropic}. At the elevated growth temperature, dislocations in the existing layers can easily move to the edge of the new island where they annihilate. As islands of different stacking grow and coalesce, new dislocation lines are formed where they meet (cf.\ Fig.\ \ref{fig:spectra}(a)). This opens the interesting possibility to engineer the dislocation network by patterning the SiC substrate before graphene growth.

Notably, we observe strain-induced domains also in monolayer EG (Fig.\ \ref{fig:nucleation}(d)) and between the bottom two layers in trilayer QFG (visible only for some energies, e.g.\ 33 eV in Supplementary Figure 2).
The prevalence of these triangular domains in all EG and QFG samples between the two bottommost layers demonstrates that stacking domains are a direct consequence of the epitaxial graphene growth and consequently are a general feature of this material system.
The resulting dislocation network explains the linear magnetoresistance observed in bilayer QFG \cite{kisslinger2015-linear} and might be an important culprit for the generally low mobility in EG and QFG \cite{speck-QFMLG}.

\begin{figure}[!ht]
	\includegraphics[width=\columnwidth]{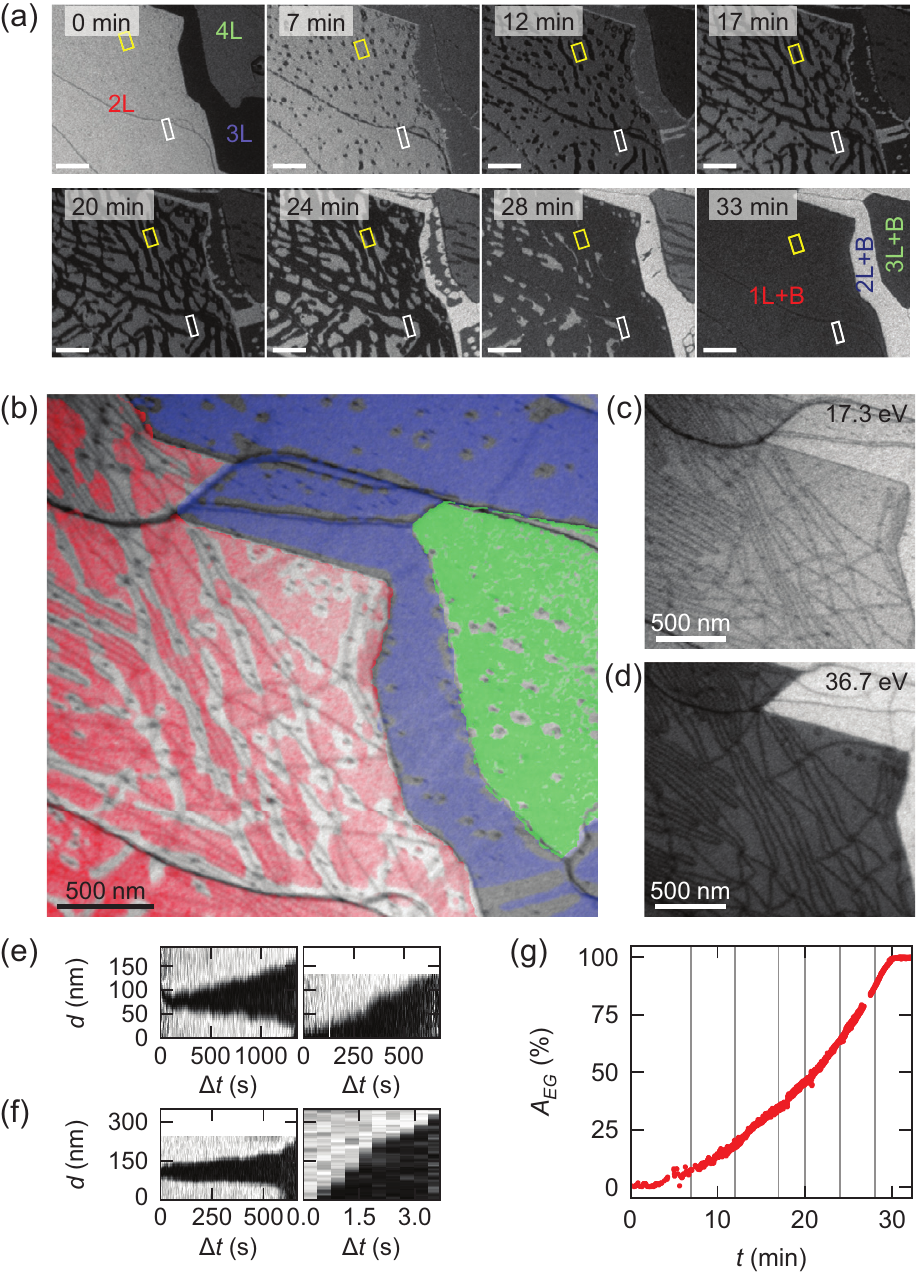}
	\caption{The hydrogen deintercalation dynamics is dominated by the graphene dislocation network.
		(a) Bright-field LEEM snapshots ($E=2.2$\,eV) of hydrogen deintercalation at \SI{\sim 1000}{\celsius} (the full time series is available as Supplementary Movie 3). Deintercalation starts in distinct points and deintercalated areas (dark in the bilayer region) grow in a strongly anisotropic fashion. Scale bars are 500\,nm.
		(b) Overlay of the deintercalation state at 15\,min with a LEEM image showing the dislocation network (dark lines) beforehand. It reveals that deintercalation proceeds faster along dislocation lines. Areas shaded in color are still intercalated, while hydrogen is already removed in the uncolored areas. 
		(c, d) Bright-field images comparing the domain boundaries before and after deintercalation, respectively. While some dislocations move slightly, the overall features remain unchanged during the process. 
		(a) to (d) show the same area as Fig.\ \ref{fig:domains}(b).
		(e) Slices along the time axis, perpendicular (left) and parallel (right) to the dislocation line marked yellow in (a), illustrate the velocity of the deintercalation front.
		(f) Same for the dislocation marked white in (a). The movement of all deintercalation fronts is roughly linear in time and much faster parallel to dislocation lines than perpendicular.
		(g) The fraction of deintercalated area $A_\text{EG}$ extracted from the bilayer area in (a) grows non-linearly in time, indicating that the process is limited by the desorption of hydrogen at the boundary between intercalated and deintercalated areas. 
	}
	\label{fig:deintercalation}
\end{figure}

The presence of these strain-induced domains in EG as well as QFG raises the question of their role during (hydrogen) intercalation. 
Since the high hydrogen pressures necessary for intercalation are not compatible with \emph{in situ} imaging, we investigate the inverse process. 
Figure \ref{fig:deintercalation}(a) shows a time series of bright-field LEEM images of the area shown in Fig.\ \ref{fig:domains}(b) recorded at \SI{\sim 1000}{\celsius}  (cf.\ Supplementary Movie 3). 
At this temperature, hydrogen slowly leaves the SiC--graphene interface \cite{riedl2009quasi, speck-QFMLG} and $n$-layer QFG is transformed back to $n-1$ layer (+ buffer layer) EG. 
The change in the reflectivity spectrum accompanied with this conversion (c.f.\ Supplementary Figure 1) yields strong contrast (e.g.\ dark in the bilayer in Fig.\ \ref{fig:deintercalation}(a)) and enables capture of the full deintercalation dynamics. 
Deintercalation starts at distinct sites where hydrogen can escape and proceeds in a highly anisotropic fashion. 
An overlay of the half deintercalated state (15\,min) with an image of the dislocations in the initial surface (Fig.\ \ref{fig:deintercalation}(b)) shows that deintercalation happens preferentially along dislocation lines. 
Although the dislocation lines are slightly mobile at higher temperatures (cf.\ Fig.\ \ref{fig:deintercalation}(c,d) before and after deintercalation, respectively), their overall direction and density is preserved during the process. 
The local deintercalation dynamics reveal details of the underlying microscopic mechanism. Figure \ref{fig:deintercalation}(e,f) show that deintercalation fronts move roughly linearly in time both perpendicular and parallel to dislocation lines. The velocity of the deintercalation fronts however, is much larger parallel to dislocation lines (up to $v_\parallel = \SI{95}{\nano\metre\per\second}$), than perpendicular to them ($v_\perp \approx \SI{0.1}{\nano\metre\per\second}$). 
This linear movement rules out that deintercalation is limited by hydrogen diffusion, but indicates that hydrogen desorption at the deintercalation front is the limiting factor. 
The non-linear growth of the fraction of deintercalated area $A_\text{EG}$ (Fig.\ \ref{fig:deintercalation}(g)) demonstrates that deintercalation is also not capped by the venting of hydrogen from the defects where deintercalation starts (7\,min in Fig.\ \ref{fig:deintercalation}(a)). 
While $v_\perp$ is the same for all areas, $v_\parallel$ varies from \SI{0.2}{\nano\metre\per\second} to \SI{95}{\nano\metre\per\second} (marked yellow and white in Fig.\ \ref{fig:deintercalation}(a), respectively), suggesting that the deintercalation process is strongly affected by the precise atomic details of the dislocations.
These findings indicate that not only the deintercalation, but also the intercalation of hydrogen and other species, which all can not penetrate graphene, is dominated by the presence of stacking domains.  
Consequently, their manipulation, e.g.\ by patterning the substrate, will open a route towards improved intercalation and tailored QFG on the wafer-scale. 

We conclude that graphene on SiC is a much richer material system than has been realized to this date.
Specifically, we show that domains of AB and AC Bernal stacking orders are always present in this material even though its layers appear perfectly crystalline to most methods. 
We deduce that these domains are formed between the two bottommost carbon layers (either graphene and buffer layer for EG or bilayer QFG) by strain relaxation. 
In addition, the nucleation of grains of different stacking order during growth causes larger domains in thicker layers. 
We show that dislocation lines between domains dominate hydrogen deintercalation dynamics, highlighting their importance for intercalation as well. 
By engineering these dislocation networks, we foresee wide implications for customized QFG for electronic applications. 
Moreover, the dislocation networks observed here can yield a wafer-scale platform for topological \cite{ju2015-TI-transport} and strongly correlated electron phenomena \cite{hunt2013-butterfly, herrero2018-magic-Mott, herrero2018-superconductivity} when tailored into periodic structures.

\begin{acknowledgments}
We thank Marcel Hesselberth and Douwe Scholma for their indispensable technical support.
This work was supported by the Netherlands Organisation for Scientific Research (NWO/OCW) via the VENI grant (680-47-447, J.J.) and as part of the Frontiers of Nanoscience program. It was supported by
the Spanish Ministry of Economy and Competitiveness MINECO, Grant No.\ FIS2016-76617-P, as well as by the DFG through SFB953.
\end{acknowledgments}


\appendix*
\section{Methods}
\paragraph*{Sample fabrication} 
Graphene growth is carried out on commercial 4H-SiC wafers (semi-insulating, nominally on axis, RCA cleaned) at $\sim$\SI{1700}{\celsius} and \SI{900}{\milli\bar} Ar pressure for $\sim$\SI{30}{\minute} as described in Ref.\  \onlinecite{emtsev2009-towards}. To convert EG to bilayer QFG via hydrogen intercalation, the sample is placed in a carbon container and heated to \SI{970}{\celsius} for \SI{90}{\minute} at ambient hydrogen pressure as described in Ref.\ \onlinecite{riedl2009quasi,speck-QFMLG}. Samples with small bilayer patches on large substrate terraces are achieved in a three-step process. First, SiC substrates are annealed at $\sim$\SI{1700}{\celsius} and \SI{900}{\milli\bar} Ar pressure for \SI{30}{\minute} in a SiC container to enable step bunching. Second, unwanted graphitic layers formed during this process are removed by annealing the sample at \SI{800}{\celsius} in an oxygen flow for \SI{30}{\minute}. Third, graphene growth is carried out as described above.

\paragraph*{Low-energy electron microscopy}
The LEEM measurements are performed using the aberration correcting \mbox{ESCHER} LEEM facility \cite{schramm2011low} which is based on a commercial SPECS P90 instrument and provides high-resolution imaging.
Limitations on the angles of the incident and imaging beams make dark-field imaging in the canonical geometry, where the diffracted beam used for imaging leaves the sample along the optical axis, impossible. 
Instead, we use a tilted geometry where the incident angle is chosen such that the specular beam and the refracted beam used for imaging leave the sample under equal, but opposite, angles (illustrated in Fig.\ \ref{fig:domains}f,i).
The tilted incidence yields an in-plane $k$-vector, which influences the reflectivity spectrum \cite{jobst2015-ARRES, jobst-ARRES-GonBN}. This is taken into account in our calculations, but needs to be considered when comparing to other LEEM and LEED data. 
Microscopy is performed below $2\cdot10^{-9}$\,mbar and at \SI{600}{\celsius}, to prevent the formation of hydrocarbon-based contaminants under the electron beam.
Images are corrected for detector-induced artifacts by subtracting a dark count image and dividing by a gain image before further analysis. Fig.\ 3 is corrected for uneven illumination by dividing by the beam profile. Additionally, the minimum intensity in images shown is set to black and maximum intensity is set to white to ensure visibility of all details. All dark-field images and images showing dislocation lines are integrated for 4\,s, all other images for 250\,ms.

\paragraph*{Computations}
All calculations were performed with a full-potential linear augmented plane
waves method based on a self-consistent crystal potential obtained within the
local density approximation, as explained in Ref.\ \onlinecite{krasovskii1999-augmented}. 
The \emph{ab initio} reflectivity spectra are obtained with the all-electron Bloch-wave-based
scattering method described in Ref.\ \onlinecite{krasovskii2004-augmented}. 
The extension of this method to stand-alone two-dimensional films of finite thickness was introduced
in Ref.\ \onlinecite{nazarov2013-scattering-2D}. 
Here, it is straightforwardly applied to the case of
finite incidence angle to represent the experimental tilted geometry. 
An absorbing optical potential $V_\mathrm{i}=\SI{0.5}{\electronvolt}$ was
introduced to account for inelastic scattering: the imaginary potential
$-iV_\mathrm{i}$ is taken to be spatially constant over a finite slab (where the
electron density is non-negligible) and to be zero in the two semi-infinite
vacuum half-spaces. In addition, a Gaussian broadening of \SI{1}{\electronvolt} is applied to account for experimental losses.

\bibliography{bilayer}

\end{document}